\documentclass[aps,showpacs,pra,twocolumn]{revtex4}
\usepackage{epsfig,amssymb,amscd}
\begin{document}
\bibliographystyle{prsty}

\title{\bf Single spin detection by qubit SWAP to a molecular nanomagnet}
\author{M. Feng $^{1,2}$ \footnote[1]{Electronic address: mfeng@thphys.may.ie} 
and J. Twamley $^{1}$ \footnote[2]{Electronic address: Jason.Twamley@MAY.IE}} 
\affiliation{$^{1}$ Department of Mathematical Physics, National University of Ireland, Maynooth, Co. Kildare, Ireland \\
$^{2}$ Wuhan Institute of Physics and Mathematics, Chinese Academy of Sciences, Wuhan, 430071, China}
\date{\today}

\begin{abstract}

Spin state detection is a key but very challenging step for any spin-based solid-state quantum computing technology. 
In fullerene based quantum computer technologies, we here propose to detect the 
single spin inside a fullerene by transferring the quantum information from the endohedral spin  to the ground states 
of a molecular nanomagnet Fe$_{8}$, with large spin S=10. 
We show how to perform the required SWAP operation and how to read out
the information through state-of-the-art techniques such as micro-SQUID.
 
\end{abstract}
\vskip 0.1cm
\maketitle

\section{introduction}

Among the variety of promising technologies to carry out quantum information processing, spin-based solid-state qubit 
technologies have great appeal through the potential use of  
large-scale fabrication techniques to scale up a quantum computer design \cite{see}. Although there have been a number of proposals
for carrying out universal quantum gates in this respect, the read out of the information from individual spin-based 
qubits - the very last and necessary step in any quantum processor routine - is still a major difficulty. 

The main idea of current theoretical  proposals for single spin detection is to convert spin-state encoded quantum 
information into charge state encodings. Since the energy of a 
multi-electron system is spin-dependent, due to the Pauli exclusion principle,  one can engineer a current whose 
spin polarization depends on spin-qubit state and thus provides a measurement of the electronic spin state
\cite {kane,loss}. Experimentally, Scanning- Tunneling- Microscope electron spin resonance (ESR) has reached the single spin 
level in the case  of iron impurities in Silicon \cite {rao}. However the precise mechanism responsible for this effect is not 
yet well understood. 
Magnetic Resonance Force Microscopy may also become a useful technique for single electron spin detection where the
cantilever oscillation is resonantly driven by a single spin \cite{berman}. However this technique will require further 
development before this technique reaches  single spin sensitivity although significant advances have been recently 
demonstrated \cite{rugar}. Recently, the movement of individual electrons has been evidenced by a  $C_{60}$-based 
single electron transistor through electro-mechanical coupling \cite {park}. It can be shown that  this arrangement can be used
as a single spin detector in the presence of an external magnetic field \cite {feng}.

Although our focus here is to use a Micro-SQUID (Superconducting Quantum Interference Device - a device capable of distinguishing 
large spin difference ($\Delta m = 30$) \cite {pakes}),
we argue that to obtain sufficient sensitivity it will be useful to combine this device with methods which could convert the small 
spin-1/2 qubit signal  
into a system with a much larger spin. What we want to detect is the $m_s$ component of the electronic spin state of the dopant 
atom inside a fullerene $C_{60}$. This doped fullerene or endohedral fullerene is the primary host for quantum information in 
a number of fullerene based quantum computer designs which include the endohedral molecules  
$N@C_{60}$ or $P@C_{60}$ \cite {harneit,suter,jason}. In
these types of designs, the quantum information is encoded in the nuclear spin of the endohedral atom. The execution of quantum 
gates utilises the hyperfine interaction   
and magnetic dipolar coupling via the application of standard  nuclear magnetic resonance (NMR) and ESR pulse sequences.
It has been clearly  demonstrated in \cite {jason} that  fullerene - based quantum computing meets the requirement for a quantum 
computer, except for the lack of an effective readout technique. Since the nuclear spin is less sensitive to the external 
environment, our 
detection of a single spin state would be carried out on the electronic spin state after the nuclear spin states have been 
swapped to the electronic spin states.

Since the problem we consider is only related to the readout of quantum information from a single electron spin state, we will neglect 
nuclear spins in this paper. 
There are three valence electrons in the dopant atom $N$ or $P$, whose ground state in the presence of magnetic field is 
$|\uparrow\uparrow\uparrow\rangle$. Because other excited states are of much higher energy, the only spin state suitable for 
encoding qubits is $|\uparrow\uparrow\uparrow\rangle$, which can be considered 
as a single spin-3/2 state with quartet components $|m_{s}=\pm 3/2\rangle$ and $|m_{s}=\pm 1/2\rangle$ in a magnetic field. It has 
been shown in \cite{jason} that universal quantum information processing can be carried out independently with inner 
qubits $|\pm 1/2\rangle$ or outer qubits $|\pm 3/2\rangle$ . Therefore what we want to detect is whether the 
inner qubit  is in  $|1/2\rangle$ or $|-1/2\rangle$ or whether the outer qubit is in  $|3/2\rangle$ or $|-3/2\rangle$.

\section {coupling system of a fullerene and a $Fe_{8}$}

Since the electrons of the dopant atoms of $C_{60}$ are always trapped in the cage made of 60 carbon atoms, the qubit 
detection mechanisms suggested in 
\cite{kane,loss}, cannot be applied here as these endohedral electrons cannot be liberated 
from the cage without destroying the delicate qubit information encoded in their $m_s$ spin states. So the key step of our scheme 
is to swap the encapsulated spin state to an outside auxiliary
spin state which is easily detectable. We consider the following Hamiltonian describes the endohedral 
electronic system coupled to a nearby auxiliary large-spin system,
\begin{equation}
H=H_{1} + H_{2} + H_{I}\;\;\label{one}
\end{equation}
where $H_{1}=-g_{1}\mu_{B}B_{z} S_{1z}$ denotes the spin (to be detected), inside the fullerene with $g_{1}=2$, $\mu_{B}$ being 
the Bohr magneton, B$_{z}$ being the magnetic field strength along z-axis, and $S_{1z}$ the Pauli matrix for $S=3/2$. $H_{2}$ is related
to the auxiliary system with large spin and $H_{I}$ is the coupling between the two systems in a magnetic field.

We employ a molecular nanomagnet made of Fe$_{8}$ \cite {exp}, which is small but has a relatively high magnetic moment, as the 
auxiliary system. It has already been shown that one can experimentally prepare a single Fe$_{8}$ crystal in the ground states 
$|\pm 10\rangle$ at very low temperatures ($< 360$ mK) \cite {wern2}. It has also been shown that one can observe  the quantum 
tunneling of magnetization in a single Fe$_{8}$ crystal, which will prove  useful for our SWAP operation. 
 The lowest energy levels of Fe$_{8}$ can be described as a spin-10 Hamiltonian \cite {wern2,loss1}
\begin{equation}
H_{2}=-DS_{2z}^{2}+ H^{tran} -g_{2}\mu_{B}B_{z} S_{2z}\;\;,\label{two}
\end{equation}
where $D \approx 0.275$ K is the axial anisotropy constant and $g_{2}\approx g_{1}$. In what follows, we define $\omega=g\mu_{B} B_{z}$ 
in units of $\hbar=1$, where $g=g_{1}\approx g_{2}$. $S_{2z}$ is the Pauli matrix for $S=10$. $H^{tran}$ is the transverse anisotropy 
due to the applied magnetic field in x-y plane, which yields the tunnel splitting \cite{garg}. The exact form of $H^{tran}$ is not 
important in our discussion. Moreover, the term concerning $S_{z}^{4}$ whose coefficient is much smaller than D is omitted in 
Eq. (\ref{two}). 
We suppose that the detected endohedral spin is coupled to the auxiliary spin of the molecular magnet by magnetic dipolar 
interaction, which is 
generally described as
$H_{I}=J_{0}(A+B+C+E+F+G)$ where $A=(1-3\cos^{2}\theta)S_{1z}\otimes S_{2z}$, $B=-(1/4)(1-3\cos^{2}\theta)(S_{1+}\otimes S_{2-}
+S_{1-}\otimes S_{2+})$, $C=-(3/2)\sin\theta\cos\theta e^{-i\phi}(S_{1z}\otimes S_{2+}+S_{1+}\otimes S_{2z})$, $E=C^{*}$, 
$F=-(3/4)\sin^{2}\theta e^{-i2\phi}S_{1+}\otimes S_{2+}$,
and $G=-(3/4)\sin^{2}\theta e^{i2\phi}S_{1-}\otimes S_{2-}$ with $\theta$ and $\phi$ the usual polar and azimuth angles, respectively.

Eq. (\ref{one}) can be exactly solved by numerical computation. For simplicity, however, we will  
investigate the system under consideration following the assumptions below: (i) The transverse anisotropy terms will be neglected 
in the following deduction as done in \cite{wern}, because it is not essential to our conclusion. (We will go beyond this 
assumption when discussing the quantum tunneling of magnetization.) (ii) Only the Ising-type term $A$, related to $S_{1z}\otimes 
S_{2z}$, 
will be retained in H$_{I}$, which is 
valid in the weak coupling limit \cite {jason}. As we will show below, in our implementation, the detuning  is much 
larger than J$_{0}$, so we can safely
omit the terms in H$_{I}$ other than $A$. As a result, the Hamiltonian under consideration can be simply rewritten as
\begin{equation}
H_{c}=-\omega (S_{1z}+ S_{2z}) -DS_{2z}^{2} + J S_{1z}\otimes S_{2z},\label{three}
\end{equation}
where $J=J_{0}(1-3\cos^{2}\theta)$ and we assume $J=0.0175$ K \cite{exp1}. Since all the terms in Eq. (\ref{three}) are diagonal, 
it is easy to obtain the eigenstates as well as the corresponding eigenenergies, as shown in Table \ref{tab1}.

\section {swap implementation}

Our scheme consists of two essential steps: the first step is to swap the detected endohedral electronic spin states inside the 
fullerene to the ground states 
$|\pm 10\rangle$ of Fe$_{8}$. Then as the second step, we measure the states $|\pm 10\rangle$ from Fe$_{8}$. Since a SWAP consists 
of three controlled-NOT 
(CNOT) operations, we will investigate in this section how to carry out CNOT within our model. To this end, we will make use of the 
quantum tunneling of magnetization in a single Fe$_{8}$ molecule.

{\it 1. Implementation of CNOT$_{21}$}.  Table \ref{tab1} shows the existence of degenerate transitions in the subspace regarding 
$H_{1}$. These transitions are
 heavily dependent on the coupling (or neighboring) spin states in $H_{2}$. Based on this characteristic, the irradiation of a ESR 
pulse with a resonant frequency  on the first spin yields, in the interaction representation, an effective Hamiltonian ($\hbar=1$) 
for the subsystem of $H_{1}$,
\begin{equation}
\tilde{H} = \Omega S_{1x},
\end{equation}  
where $\Omega$ is the Rabi frequency, 
$$S_{1x}=\frac {1}{2}\pmatrix {0 & \sqrt{3} & 0 & 0 \cr \sqrt{3} & 0 & 2 & 0 \cr 0 & 2 & 0 & \sqrt{3} \cr  0 & 0 & \sqrt{3} & 0},$$
and the resonant frequency is one of the degenerate transition frequencies labeled in Table \ref{tab1}.
For a $\pi$ pulse irradiation of ESR, i.e. $\Omega t=\pi$, $H_{I}$ yields the operator
$$\hat{P}=i\pmatrix {0 & 0 & 0 & 1 \cr 0 & 0 & 1 & 0 \cr 0 & 1 & 0 & 0 \cr  1 & 0 & 0 & 0},$$ which works independently in the 
subspace spanned by $|\pm 3/2\rangle$ or the one spanned by $|\pm 1/2\rangle$.

Therefore, with the operator $\hat{P}$, we can flip states $|\pm 3/2\rangle$ or $|\pm 1/2\rangle$ of a single qubit with an ESR pulse 
whose frequency is determined by the neighboring spin state in $H_{2}$. That is actually a CNOT$_{21}$ operation. 
Theoretically, the fidelity of this selective pulse method depends on more detailed characteristics  of the physically coupled systems. 
The implementation time is determined by the Rabi frequency $\Omega$.  

{\it 2. Implementation of CNOT$_{12}$}. Due to the term $S_{2z}^{2}$,  the levels in Fe$_{8}$ are not equidistant and so
CNOT$_{12}$ cannot be carried out by the above method with selective ESR pulses. Let us simply consider the case of very low 
temperature, in 
which only the ground states $|\pm 10\rangle$ of $H_{2}$ are populated. By considering $|\pm 10\rangle$ coupled with the 
possible qubit states of H$_{1}$, we show the magnetic field dependence of the energy in Fig. 1, where the displayed  crossing points 
correspond to the doublet degenerate cases. However, if 
we include the neglected term $H^{tran}$ in our calculation, all of these crossing points would turn into avoided crossings due to 
the tunneling splitting. So by sweeping the magnetic field B$_{z}$ through these (avoided) crossing points, 
we should have tunneling between different magnetization states. For example, with the magnetic field B$_{z}$ swept through 0.019 T 
(from 0.019$_{-}$ T to 0.019$_{+}$ T), we have the magnetization tunneling from $|3/2, -10\rangle$ to $|3/2, 10\rangle$.  Note that 
in Fig. 1 the (avoided) crossing points of different kinds of lines mean the places where the magnetization tunneling occurs with 
very small probability due to the second or higher-order process, e.g. $|1/2, 10\rangle\rightarrow |3/2, -10\rangle$.
Moreover, since the eigenenergies associated with $|n, m \rangle,\;\;n=\pm 1/2, \pm 3/2$ and $m= \pm 9, \pm 8,...$ are much higher, 
and thus are physically hard to reach in the low temperature  case, we omit them in Fig. 1.  

The fidelity of our scheme is affected by the following: (1) We must have precise knowledge of the dipolar coupling 
strength $J$, as this determines the quality of the CNOT gates we perform. The $J-$value can be obtained experimentally by 
interrogative ESR pulses. 
(2) We must have very small linewidth selective ESR pulses. 
 Since the wavelength of the ESR pulse is much larger than the distance between Fe$_{8}$ and C60, for avoiding exciting
Fe$_{8}$ when we perform CNOT$_{21}$, it is required that the linewidths of ESR pulses be narrower than $J$, the minimal difference 
between degenerate transition frequencies. (3) We must accurately control the tunneling time in 
the performance of CNOT$_{12}$.  This time depends on the sweep rate of the magnetic field and the tunnel splitting. We prefer
a fast tunneling to make the implementation time of SWAP shorter than the decoherence time. In this case, we
need a large tunnel splitting at the avoided crossing points.

\section {detection of states $|\pm 10\rangle$}

Due to the very high sensitivity (which reaches $10^{-16}$ electromagnetic units \cite{wern1}), and from the full spectrum analysis 
we have done for nanomagnets \cite {barra}, the micro-SQUID can  hopefully be used to 
measure the spin states of single Fe$_{8}$ crystals directly by standard spectroscopy with pulsed ESR \cite{loss1}. 

\section {discussion and conclusion}

Although the experimental value of the coupling $J$ is not yet known, we expect $J$ to be suitably large for our purpose 
in implementing the CNOT gates. First, as mentioned above, the prerequisite of a perfect CNOT$_{21}$ 
implementation is that the linewidth of the ESR pulse is smaller than $J$. So the larger the coupling $J$, the less strict the 
requirement 
for the linewidth of the ESR pulse. Secondly, a larger coupling $J$ is advantageous to achieve more accurate implementations of 
the magnetization 
tunneling. In terms of our numerical calculation, if $J$ is very close to zero, all the (avoided) crossing points
would be nearly overlapping. It implies that our scheme would not work because we could not distinguish different qubit spin states
from the tunneling signal. On the other hand, $J$ cannot be too large. To keep the weak coupling limit in Eq. (\ref{three}) valid, 
$|J|$ should be much smaller than min \{$|\omega|$, D\}. In our case with  $J=0.0175$ K corresponding to 350 MHz, 
ESR pulses with much narrower linewidths have already been achieved experimentally \cite {meyer}.For the magnetization 
tunneling $|3/2, -10\rangle \leftrightarrow |3/2, 10\rangle$  occurring at $B_{z}=0.019$ T, $|J| \sim |\omega|/2 \ll D$. Nevertheless,
as long as the dipolar angle $\theta$ in $H_{I}$ is close to $\pi/2$, terms $C$ and $E$  in H$_{I}$ would be nearly zero, and other 
terms (except $A$) could be neglected in the weak limit and so Eq. (\ref{three}) still holds. 

An essential point for implementing our scheme is that the SWAP time should be shorter than the decoherence time of the system. For 
Fe$_{8}$, the T$_{1}$ times for the 
$|\pm 10\rangle$  ground states are very long, while the encapsulated spin states in the 
fullerene also posses  long T$_{1}$ times,  ~1 sec at 7 K \cite {knorr}, and probably several seconds for lower temperatures 
\cite {meyer}. Thus the dominant source of decoherence in our model would be due to the hyperfine level broadening produced by the 
nuclei \cite {wern2, sessoli1}. This causes undesired dephasing in our scheme during the quantum tunneling of magnetization.
We consider the SWAP time to be $2\pi/\Omega + $ T$_{0}$ with T$_{0}$ being the magnetization tunneling time. With current experimental 
techniques, $\Omega= 20 \sim 30$ MHz is available \cite {meyer}, and T$_{0}$ can be from nanosec to sec depending on the transverse 
magnetic field \cite {wern100}. In the experiment of \cite {wern3}, it is shown that,
in temperatures below 350 mK, and in the presence of B$_{z}=42$ mT and a transverse B-field ~200 mT, the line broadening is
about 0.8 mT, corresponding to 22.4 MHz in units of frequency. So if we assume $\Omega =30$ MHz, to carry out our scheme, 
we require that T$_{0}$ be shorter than 71 nanosec.

Another point we should mention is the initial state preparation of Fe$_{8}$. Since the detected spin inside the fullerene in 
the readout stage should be well polarized, we can simply convert the spin information from $H_{1}$ to $H_{2}$ by performing 
CNOT$_{12}$, instead of SWAP. This would simplify the readout scheme we mentioned above. To this end, however, we need to 
precisely prepare the initial states
of Fe$_{8}$ to be $|10\rangle$ or $-|10\rangle$. This too can be done by the quantum tunneling of magnetization. When we sweep the 
B$_{z}$ field quickly, we can have an oscillation between $|10\rangle$ and $|-10\rangle$, whose frequency is 
heavily related to the tunnel splitting. By stopping the sweep field at an exactly chosen time, we can have a perfect initial state
$|10\rangle$ or $|-10\rangle$ of Fe$_{8}$.

In summary, we have proposed a potential method to efficiently detect the single spin state inside the fullerene by means of an
auxiliary large spin nanomagnet. Since the spin states of the large-spin nanomagnet are measurable with current experimental 
techniques (e.g. Micro-SQUID), we can reliably detect the qubit-encoded spin states inside the fullerene by swapping the detected 
state into the ground states of a nanomagnet.

\section {acknowledgment}

The authors acknowledge thankfully the discussion with Wolfgang Wernsdorfer, Derek Mc Hugh and Graham Kells. MF is grateful for 
support from Chinese Academy of Sciences. The work is supported by EU Research Project QIPDDF-ROSES under contract number 
IST-2001-37150.

\newpage
\begin{table}
\centering
\caption {The eigenstates and the corresponding eigenenergies obtained from Eq. (\ref{three}), where the degenerate
transition frequencies between two nearest neighbor rows in the same column are listed in the last row. Due to the axial
anisotropy term, there is no degenerate transition between two nearest-neighbor columns. See text for the details.}
\bigskip
\begin{tabular}{c|c|c|c|c}
\hline
$|\frac {3}{2}, 10 \rangle$ & $|\frac {3}{2}, 9 \rangle$ & $\cdots\cdots$ &
$|\frac {3}{2}, -9 \rangle$ & $|\frac {3}{2}, -10 \rangle$ \\ 
\hline
$-11.5\omega - 100 D + 15J$ & $-10.5\omega - 81 D + 13.5J$ & $\cdots\cdots$ &
$7.5\omega - 81 D - 13.5J$ & $8.5\omega - 100 D - 15J$ \\ 
\hline
$|\frac {1}{2}, 10 \rangle$ & $|\frac {1}{2}, 9 \rangle $ & $\cdots\cdots$ &
$|\frac {1}{2}, -9 \rangle$ & $|\frac {1}{2}, -10 \rangle,$ \\
\hline
$-10.5\omega - 100 D + 5J$ & $-9.5\omega - 81 D + 4.5J$ & $\cdots\cdots$ &
$8.5\omega - 81 D - 4.5J$ & $9.5\omega - 100 D - 5J$ \\ 
\hline
$|-\frac {1}{2}, 10 \rangle$ & $|-\frac {1}{2}, 9 \rangle $ & $\cdots\cdots$ &
$|-\frac {1}{2}, -9 \rangle$ & $|-\frac {1}{2}, -10 \rangle,$ \\
\hline
$-9.5\omega - 100 D - 5J$ & $-8.5\omega - 81 D - 4.5J$ & $\cdots\cdots$ &
$9.5\omega - 81 D + 4.5J$ & $10.5\omega - 100 D + 5J$ \\ 
\hline
$|-\frac {3}{2}, 10 \rangle$ & $|-\frac {3}{2}, 9 \rangle$ & $\cdots\cdots$ &
$|-\frac {3}{2}, -9 \rangle$ & $|-\frac {3}{2}, -10 \rangle$ \\ 
\hline
$-8.5\omega - 100 D - 15J$ & $-7.5\omega - 81 D - 13.5J$ & $\cdots\cdots$ &
$10.5\omega - 81 D + 13.5J$ & $11.5\omega - 100 D + 15J$ \\ 
\hline
\hline
$-\omega + 10 J$ & $-\omega + 9J$ & $\cdots\cdots$ &
$-\omega - 9J$ & $-\omega - 10 J$ \\ 
\hline\label{tab1}
\end{tabular}
\end{table}
\newpage
\begin{figure}[p]
\begin{center}
\setlength{\unitlength}{1cm}
\begin{picture}(4,15)
\put(-4,3){\includegraphics[width=12cm]{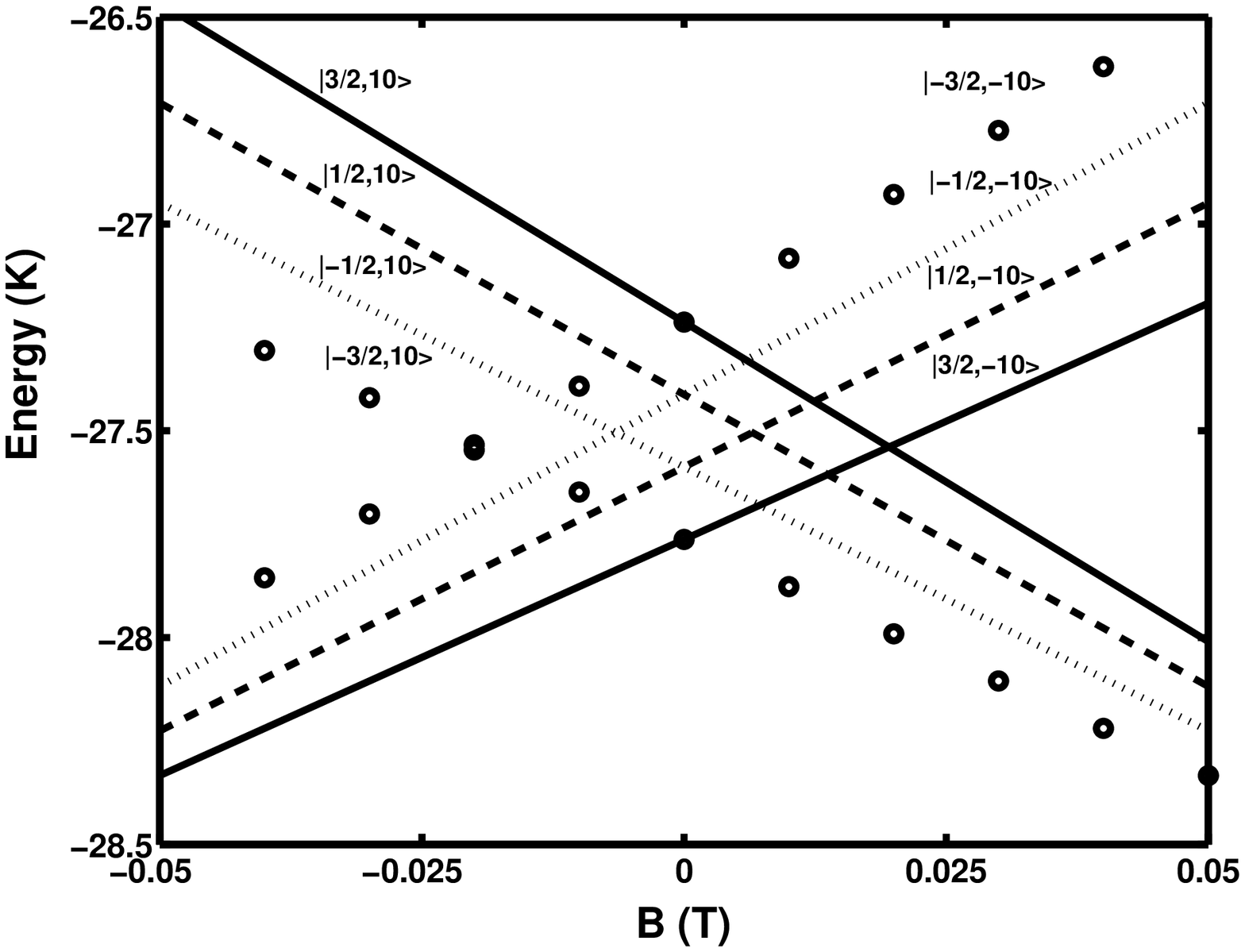}}
\end{picture}
\end{center}
\caption{Energy versus B$_{z}$ field plot for the low-lying states at very low temperature based on Eq. (\ref{three}), 
where $D=0.275 K$ and 
$J=0.0175 K$. The quantum tunneling of magnetization happens at (avoided) crossing points of the same kinds of lines. The (avoided)
crossing points of different kinds of lines corresponds to the magnetization tunneling with smaller probability due to second or higher 
order process.}
\label{Fig1}
\end{figure}


\begin{thebibliography}{22}
\bibitem{exp} We can also employ the molecular nanomagnet Mn$_{12}$ which has a large total spin $S=10$. But as shown in 
\cite {wern2}, even in the temperature below 60 mK, the excited states of Mn$_{12}$ are still populated. This would affect our scheme. 
So we choose Fe$_{8}$. But since Mn$_{12}$ and Fe$_{8}$ have many common characters, we will also cite some papers for Mn$_{12}$
in our discussion.
\bibitem{exp1} Since there is no experimental data for that, this value is theoretically supposed due to the following fact: If a 
nanomagnet Fe$_{8}$ with the same size as a fullerene C$_{60}$ ($<$ 0.6 nm) is put next to a C$_{60}$ by a spin-spin distance of 1 nm, 
the dipolar coupling between them should be similar to that between two fullerenes \cite{harneit} with the coupling strength 50 MHz.
Considering the larger spin of the Fe$_{8}$, we assume the dipolar coupling here is larger than in \cite{harneit} by 7 times.
\end{thebibliography}
\end{document}